# Site effects in an alpine valley with strong velocity gradient: interest and limitations of the 'classical' BEM

Nicolas Delépine[1], Jean-François Semblat[2]
[1] Université Paris-Est, IFSTTAR, Paris, France
(now at IFP Energies nouvelles, France, nicolas.delepine@ifpen.fr)
[2] Université Paris-Est, IFSTTAR, Dept Geotech.Eng; Environment and Risks, Paris, France, jean-francois.semblat@ifsttar.fr

**ABSTRACT** – Seismic waves may be strongly amplified in deep alluvial basins due to the velocity contrast (or velocity gradient) between the various layers as well as the basin edge effects. In this work, the seismic ground motion in a deep alpine valley (Grenoble basin, French Alps) is investigated through various 'classical' Boundary Element models. This deep valley has a peculiar geometry ("Y"-shaped) and involves a strong velocity gradient between surface geological structures. In the framework of a numerical benchmark [21-23], a representative cross section of the valley has been proposed to investigate 2D site effects through various numerical methods. The 'classical' Boundary Element Method is considered herein to model the strong velocity gradient with a 2D piecewise homogeneous medium.
For a large incidence angle, the transfer functions estimated from plane SH waves are close to the one computed with shallow SH point sources. The fundamental frequency is estimated at 0.33 Hz (SH wave) and the agreement with previous experimental results (spectral ratios) is good. Comparisons between 1D and 2D amplification are then performed: the values of the fundamental frequency and the corresponding amplitude are larger in 2D. Converting frequency domain results into the time domain, we underline surface waves generated at the valley edges and the directivity effect for the amplified wave-field. In the time domain for plane SV-wave, we also computed the ground motion for a strong seismic event (M=6): time duration and peak ground velocity are found to be 3 times larger than for the input signal. Such 2D models involving basin effects are then capable to recover the high amplification level measured in the field. Nevertheless, to deal with complex 3D basins as also proposed in the "ESG" benchmark [21-23], the capabilities of the classical Boundary Element Method are limited. As shown recently [43,47,52], such improvements as the fast multipole formulation (FM-BEM) may be a promising alternative for future 3D simulations.

## 1. Introduction

The local geological structure of a site can strongly modify the seismic wave propagation and it can lead to large amplifications and strong spatial variations of the ground motion [1-3]. The amplification process may enhance the impact of earthquakes even for moderate ones.

Both experimental and numerical approaches allow the characterization of site effects. Various numerical methods have been considered for 2D [4-13], 2.5D [14] and 3D analyses [15-19]. For such models, the need for reliable field data is then strong since the amplification process is very sensitive to the properties and geometry of the geological layers.

In the present analysis, we consider a deep alpine valley (Grenoble basin, French Alps). This deep valley has a peculiar geometry ("Y"-shaped) and involves a strong velocity gradient in the surface geological structures. The 500m deep Grenoble basin is schematized in Fig.1 and several acceleration recordings of the Laffrey 1999 earthquake are displayed [20]. The reference bedrock site is called *OGMU* (top left) and the other stations are located at the surface of the deposit. These stations are part of the French accelerometric network ("RAP", http://www-rap.obs.ujf-grenoble.fr/). As shown in this figure, the Grenoble basin strongly amplifies the seismic motion due to multiple reflections and diffractions at the basin edges.

In the framework of an international numerical benchmark [21-23], a 3D model and a representative 2D cross section of the valley have been proposed to investigate site effects through various numerical methods. The benchmark proposed an idealized velocity profile in the alluvial deposit (strong velocity gradient). In this paper, since the classical Boundary Element Method is limited to piecewise homogeneous media and has a significant computational burden in 3D [24], we shall model the 2D geological profile proposed in the benchmark. The perspectives in terms of 3D BEM modeling will be also discussed in the following.

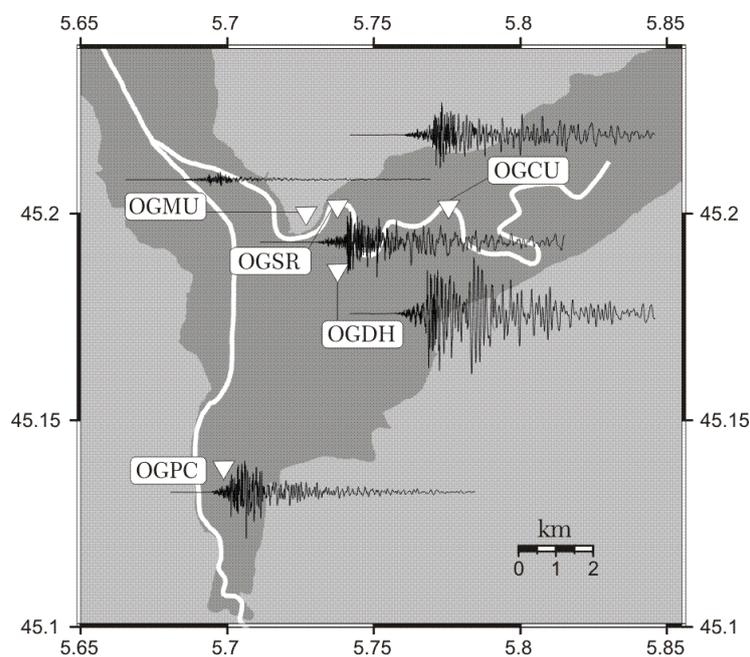

***Fig. 1.*** *View of the alpine valley (Grenoble basin) and velocity recordings at the surface for the Laffrey 1999 earthquake [20].*

## 2. Modeling seismic wave propagation by the BEM
### 2.1 *Numerical methods for wave propagation*
To analyze seismic wave propagation in 2D or 3D geological structures, various numerical methods are available:
- the finite difference method is accurate in elastodynamics but is mainly adapted to simple geometries [25-26],
- the finite element method is efficient to deal with complex geometries and numerous heterogeneities (even for inelastic constitutive models [27-29]) but has several drawbacks such as numerical dispersion and spurious wave reflections [30-37]. It may (consequently) lead to huge numerical costs in 3D elastodynamics,
- the spectral element method has been increasingly considered to analyse 2D/3D wave propagation in linear media with a good accuracy due to its spectral convergence properties [38-40] but the spurious wave reflections still have to be removed [30-37],
- the boundary element method allows a very good description of the radiation conditions but is preferably dedicated to weak heterogeneities and linear constitutive models [11,17,24,33,41,46]. It is thus difficult to deal with strong heterogeneities in 3D except if considering the original developments recently proposed to reduce the computational cost of the method [43-44,47],
- the Aki-Larner method takes advantage of the frequency-wavenumber decomposition but is limited to simple geometries [45,46],
- the scaled boundary finite element method is a kind of solution-less boundary element method [48],
- other methods such as series expansions of wave functions [49].

Furthermore, when dealing with wave propagation in unbounded domains, many of these numerical methods raise the need for absorbing boundary conditions to avoid spurious reflections [30-37]. It is also possible to couple FEM and BEM [14,24,50] allowing an accurate description of the near field (FEM model including complex geometries, numerous heterogeneities and nonlinear constitutive laws) and a reliable estimation of the far-field (BEM involving accurate radiation conditions).

### 2.2  The 'classical' Boundary Element Method
The Boundary Element Method arises from the application of Maxwell-Betti reciprocity theorem leading to the expression of the displacement field inside the domain $\Omega$ from the displacements and tractions along the boundary $\partial\Omega$ of the domain [24]. The main advantage of the method is to avoid artificial truncation of the domain in the case of infinite medium [24]. For dynamic problems, this truncation leads to artificial wave reflections giving a numerical error in the solution.

### 2.3  Elastodynamics
We consider an elastic, homogeneous and isotropic solid of volume $\Omega$ and external surface $\partial\Omega$. In this medium, the equation of motion can be written under the following form:

$$(\lambda + \mu)\nabla(\nabla \boldsymbol{u}) + \mu\Delta \boldsymbol{u} + \rho \boldsymbol{f} = \rho \ddot{\boldsymbol{u}} \tag{1}$$

where $\boldsymbol{u}$ is the displacement field, $\boldsymbol{f}$ a density of body force and $\lambda,\mu$ the Lamé parameters.

In this article, the problem is solved in frequency domain by superposition of solutions having a harmonic dependence in time of circular frequency $\omega$. The equation of motion for a steady state ($\boldsymbol{u}(\boldsymbol{x})$, $\sigma(\boldsymbol{x})$) can then be written as follows:

$$(\lambda + \mu)\nabla(\nabla \boldsymbol{u}(\boldsymbol{x})) + \mu\Delta \boldsymbol{u}(\boldsymbol{x}) + \rho \boldsymbol{f}(\boldsymbol{x}) = -\rho\omega^2 \boldsymbol{u}(\boldsymbol{x}) \tag{2}$$

This equation is written in the framework of linear elasticity but, since the analysis is performed in the frequency domain, damped mechanical properties may be considered through the complex modulus of the medium [33, 47] (it is discussed in the following).

### 2.4  Integral formulation
The integral formulation is obtained through the application of the reciprocity theorem between the elastodynamic state ($\boldsymbol{u}(\boldsymbol{x})$, $\sigma(\boldsymbol{x})$) and the fundamental solutions of a reference problem called Green kernels [24]. The reference problem generally corresponds to the infinite full space case in which a concentrated body force at point $\boldsymbol{y}$ acts in any given direction $\boldsymbol{e}$.

In the harmonic case, the Green kernel of the infinite medium corresponds to a body force field such as:

$$\rho \boldsymbol{f}(\boldsymbol{x}) = \delta(\boldsymbol{x} - \boldsymbol{y})\boldsymbol{e} \tag{3}$$

In this article, the model involves the Green functions of an infinite medium [24] or semi-infinite medium (in the case of SH-waves). The Green kernel is denoted $U_{ij}^{\omega}(x,y)$ and characterizes the complex displacement in direction $j$ at point $x$ due to a unit force concentrated at point $y$ along direction $i$. The corresponding traction for a surface of normal vector $n(x)$ is denoted $T_{ij}^{(n)\omega}(x,y)$. The application of the reciprocity theorem between the elastodynamic state ($u(x)$, $\sigma(x)$) and that defined by the Green kernel $U_{ij}^{\omega}(x,y)$ gives the following integral representation:

$$I_{\Omega}(y)u_i(y) = \int_{\partial\Omega} \left( U_{ij}^{\omega}(x,y)t_j^{(n)}(x) - T_{ij}^{(n)\omega}(x,y)u_j(x) \right) ds(x) + \int_{\Omega} \rho U_{ij}^{\omega}(x,y)f_j(x) dv(x) \quad (4)$$

where $I_{\Omega}(y)$ is 1 when $y \in \Omega$ and 0 when $y \notin \Omega$.

Numerical solution of equation (4) can be performed by collocation method or by an integral variational approach [24]. When the domain $\Omega$ is infinite and there is no source at infinite distance, it is necessary to give restrictive conditions on the behaviour of the displacement field $u(x,t)$ at infinity. These assumptions are called outgoing Sommerfeld radiation conditions. When there are some sources at infinity (denoted by the field $u^{inc}$), the Sommerfeld conditions are applied to the scattered displacement field $u^s = u - u^{inc}$.

### 2.5 Boundary integral equation and discretization

The integral representation defined by equation (4) is generally not valid for $x \in \partial\Omega$. The formulation of the boundary integral equation along $\partial\Omega$ is then not very easy to obtain as the Green kernels have singular values when $x \in \partial\Omega$. It is then necessary to regularize expression (4) to write the boundary integral equation [17,24,41]. Afterwards, the regularized solution of equation (4) is estimated by classical boundary finite elements discretization and then by collocation method, that is application of the integral equation at each node of the mesh.

With the classical BEM formulation, it is necessary to compute all the contributions between the source points and the observation points in the mesh. The computational burden may thus be large especially in complex 3D media. Recent improvements of the BEM [43,47] allow to reduce such a large numerical cost but the reduction is not sufficient for the deep alluvial basin considered herein since it has a strong velocity gradient.

We will thus consider two dimensional piecewise homogeneous models (plane or anti-plane strains). Two dimensional Green kernels of the infinite space are written using Hankel's functions [24].

### 3. BEM Model of the deep alluvial basin

As proposed in the numerical benchmark for this basin [21-22], and due to the limitations of the classical BEM to model complex 3D media, we choose the 2D geological profile with a strong velocity gradient. It is located in the north eastern part of the valley (Fig.2, top) and, as shown by the symbols in this area, detailed geophysical surveys were performed to characterize this geological profile. The simulations are performed in the frequency domain considering the classical BEM (FEM/BEM code CESAR-LCPC [42]); time domain ground motions may be post-processed from frequency domain results in a second step.

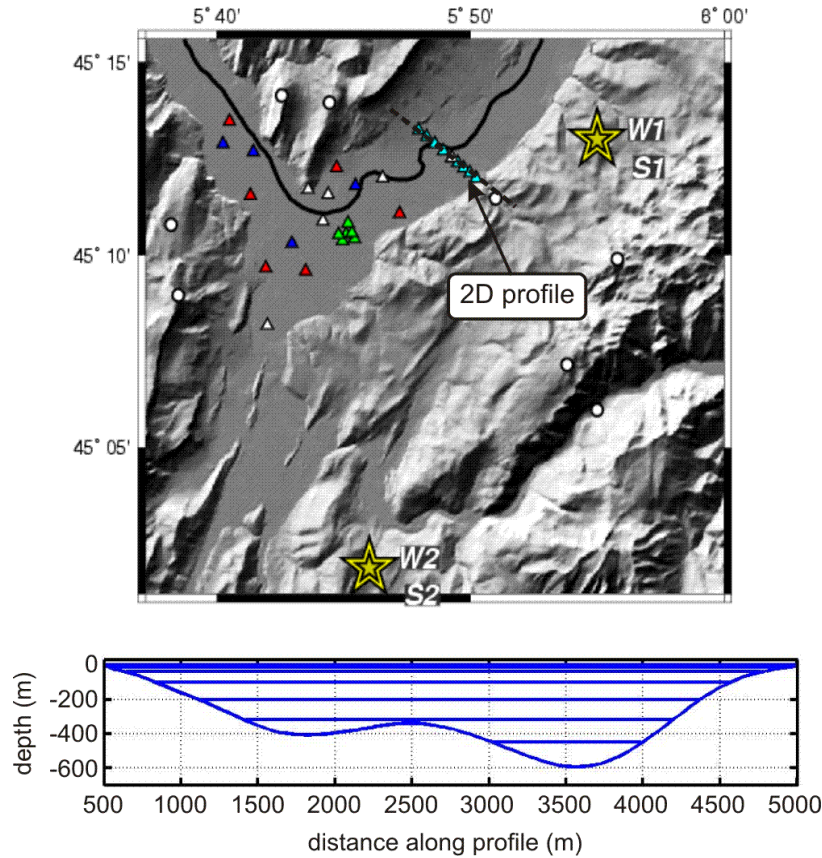

**Fig. 2.** *The Grenoble valley (top) with the locations of the 2D geological profile and of the epicenters of scenarios 1 and 2; Boundary Element model (bottom) with its 7 layers.*

### 3.1   Modeling the incident seismic motion

As depicted in Fig.2 (top), two scenarios of earthquakes were proposed in the benchmark: a first one having its epicenter in the north-eastern part of the model and the second one located in the southern part of the model. As mentioned in Fig.2 (top), weak ("*W*") and strong ("*S*") events may be considered in the (1D/2D or 3D) simulations. In the BEM model, the first scenario (local event) will be modeled through a point source and the second scenario (regional event) will be modeled through plane waves with various incidences. In all cases, monochromatic waves are considered for frequency domain simulations.

### 3.2   Modeling the velocity variations

As proposed in the benchmark [21-22], the wave velocities are supposed to increase continuously with depth (strong velocity gradient). Since the Boundary Element method requires piecewise constant velocities through several homogeneous sub-domains [24], we have divided the alluvial basin in 7 different sedimentary layers (Fig.2, bottom). The constant mechanical properties of each layer are given in Table I and plotted in Fig.3. The thickness of the surficial layers is smaller due to their lower velocity (shorter wavelength). It is not currently possible to deal with such a complex 3D model including a strong velocity gradient with the BEM. Some ongoing research on accelerated BEM techniques may allow to do so in the near future [43,47,52].

The velocity-depth variations suggested in the benchmark [21-22] are displayed in Fig. 3 (dotted lines). The piecewise constant velocities chosen for our classical BEM model (solid lines) are close to the previous reference ones (difference with the proposed gradient within a few percents).

**Table I.** Constant velocity values in the 7 sedimentary layers.

| Layer | $Z_1$ (m) | $Z_2$ (m) | $V_P$ (m/s) | $V_S$ (m/s) |
|---|---|---|---|---|
| 1 | 0 | 15 | 1459 | 337 |
| 2 | 15 | 40 | 1483 | 397 |
| 3 | 40 | 100 | 1534 | 455 |
| 4 | 100 | 200 | 1630 | 529 |
| 5 | 200 | 320 | 1762 | 604 |
| 6 | 320 | 450 | 1912 | 671 |
| 7 | 450 | 591 | 2074 | 732 |

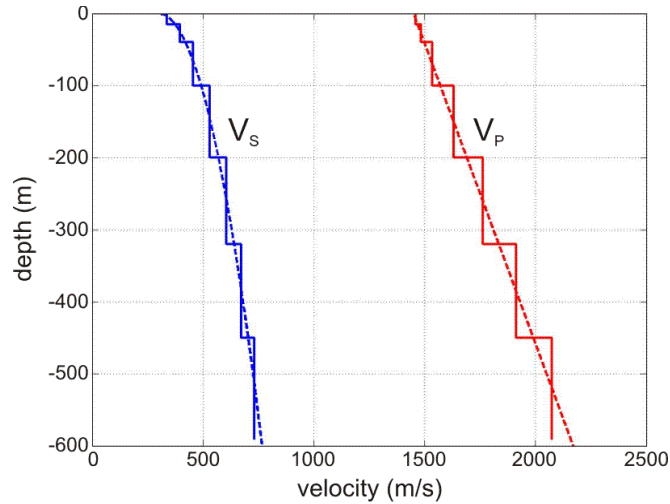

**Fig. 3.** *Velocity profile (left) for P ($V_P$ : right) and S waves ($V_S$ : left); dotted lines: benchmark values, solid lines: piecewise constant BEM model values.*

### 3.3 Modeling of damping

The boundary element model is supposed to have a linear viscoelastic behaviour. The formulation of damping considered herein corresponds to a Zener model or standard solid [33]. Since we perform the analysis in the frequency domain, we may use various other types of damping-frequency dependence [47].

This rheological model is depicted in Fig. 4 for shear response and the expression of the inverse of the quality factor $Q^{-1}$ (i.e. attenuation) is given as a function of frequency. The variations of $Q^{-1}$ with frequency are also drawn in this figure showing a peak corresponding to the maximum value of attenuation. Considering this rheological model, the complex shear modulus of the medium $\mu^* = \mu_R + i.\mu_I$ can be written as a function of frequency, short term (instantaneous) $\mu_{st}$ and long term $\mu_{lt}$ shear moduli as follows :

$$\begin{cases} \mu_R = \dfrac{\mu_{lt}\mu_{st}^4 + \omega^2\zeta^2 \mu_{st}(\mu_{st} - \mu_{lt})^2}{\mu_{st}^4 + \omega^2\zeta^2(\mu_{st} - \mu_{lt})^2} \\ \mu_I = \dfrac{\omega\zeta\mu_{st}^2(\mu_{st} - \mu_{lt})^2}{\mu_{st}^4 + \omega^2\zeta^2(\mu_{st} - \mu_{lt})^2} \end{cases} \quad (5)$$

where $\zeta$ is the viscosity coefficient of the Zener's model (Fig. 4), $\mu_R$ and $\mu_I$ are the real and imaginary parts of the complex shear modulus and the long term shear modulus is such as (see Fig. 4, for the definition of $\mu'$) :

$$\frac{1}{\mu_{lt}} = \frac{1}{\mu_{st}} + \frac{1}{\mu'}$$

As suggested in the numerical benchmark, we have taken the following values for the quality factors: $Q_S=50$ and $Q_P = \dfrac{3Q_S}{4}\dfrac{V_P^2}{V_S^2}$.

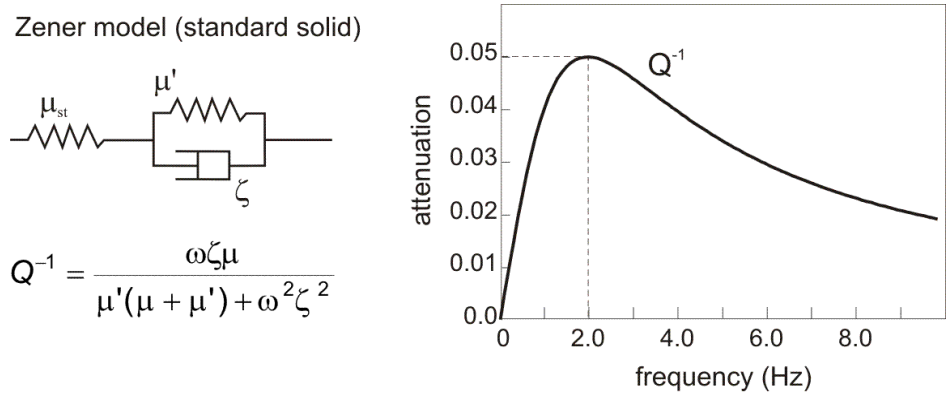

**Fig. 4.** *Zener's model (standard solid) and corresponding attenuation vs frequency dependency.*

## 4. Incident SH-wave
### 4.1. Plane SH-wave
First, various plane waves with various incidences will be considered for the second scenario (regional event).

4.1.1. Amplification in the frequency domain

Considering a vertical plane SH wave in the BEM model (incidence 90° from the horizontal), the transfer functions (Fig. 5, top left) of the surface ground motion show several amplitude maxima for frequencies ranging from 0.325 to 0.95 Hz (with a nearly constant frequency step of 0.105 Hz). They correspond to the various shear modes of the 2D basin. For increasing frequencies, the shallower parts of the basin lead to larger ground motion levels.

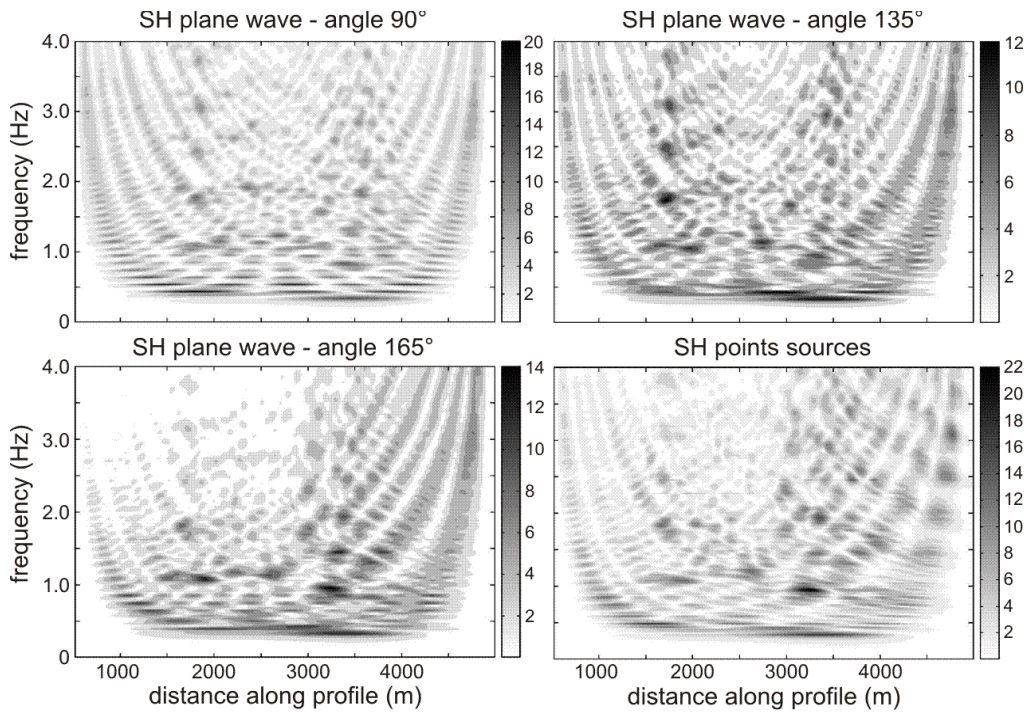

**Fig. 5.** Transfer functions from 0 to 4 Hz for a plane SH-wave and various incidence angles.

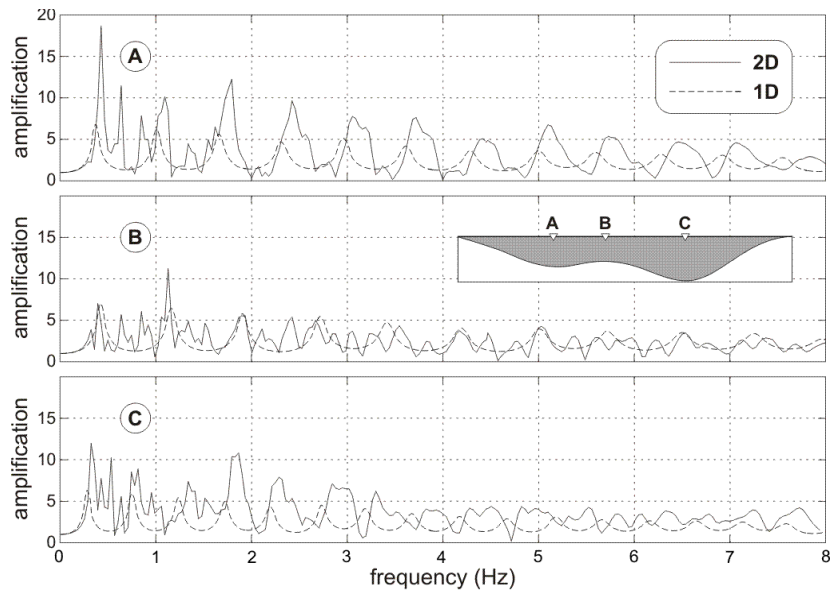

**Fig 6.** *1D and 2D transfer functions at points A, B and C (vertical plane SH-wave).*

We will now consider some specific areas along the basin surface (Fig. 6): point A is located at the center of the left part of the basin, point C at the center of the deepest area (right part) and point B in between the previous points (shallower area at the center of the whole profile).

Near point A for instance, the horizontal motion reaches nearly 20 times the amplitude of incident waves at some receivers (Figs 5 & 6). The corresponding frequency (f=0.45Hz) is higher than the fundamental frequency since this area is located in the shallower left part of the basin. The equivalent 1D model (Fig. 6, dotted line) gives an amplification value of 7 at a lower frequency value of 0.37 Hz. For higher frequency modes, the same phenomena can be observed.

For the deepest part of the basin (point C), results are similar to those obtained at point A: the discrepancy between 2D and 1D amplifications is large. However, the amplification level is significantly lower at point C when compared to point A.

At the center of the whole basin (point B), the 2D and 1D amplification levels are very close. These results show that the lateral propagation and focusing effects play a crucial role at the basin edges for both left and right part of the valley (points A & C). Whereas, at the center of the basin, the 2D and 1D transfer functions are much closer (point B).

4.1.2. Influence of the incidence angle
As already observed in previous works [1,11], the fundamental frequency value of the basin (f=0.33Hz) is not influenced by the incidence angle (Fig. 5, top right and bottom left). The amplitude at the fundamental frequency is nevertheless maximum for the vertical incidence. For a seismic excitation with a higher incidence angle (propagating from the east), larger amplifications are computed in the eastern part of the basin showing strong directivity effects (Fig. 5).

4.1.3. Time domain amplification of the seismic motion.
To analyze site effects in the basin through time domain motion amplification, we considered several Ricker wavelets at various frequencies. We combine the transfer functions in the frequency domain (Fig. 5) with the Fourier transform of the Ricker wavelets to compute the time domain ground motion for various points along the free surface.

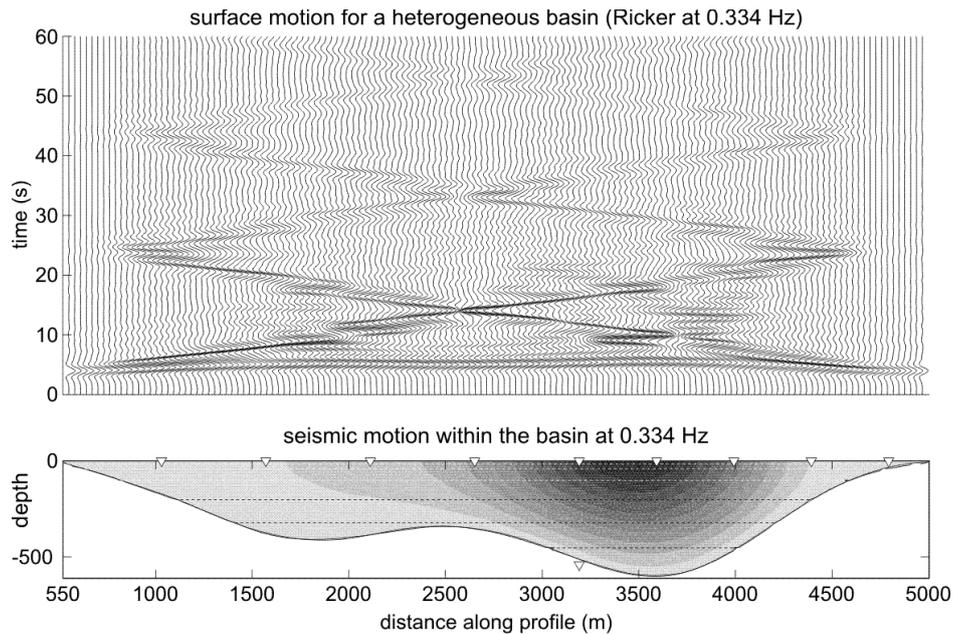

**Fig. 7.** *Top: synthetic seismogram for a vertically incident SH-wave (Ricker signal at 0.334Hz). Bottom: ground motion amplification within the basin at a given frequency (f=0.334Hz).*

We firstly consider a Ricker wavelet at a frequency corresponding to the fundamental mode of the valley (f=0.33Hz). The harmonic response of the basin is recalled in Fig. 7, bottom (the maximum spectral amplification reaches 24.5 in the darkest area).

The time domain results displayed in Fig. 7 (top) lead to the following comments:
- Around t=5s, we obtain the 1D response of the basin with different arrival times along the profile (various mean velocities). On the right part of the basin, the late arrival times obviously correspond to receivers above the deepest part of the basin.
- Until t=8s, we mainly observe the vertically reflected waves (1D effect). They lead to a significant amplification of the Ricker wavelet. The two convex parts of the basin give nearly independent responses to the seismic motion. At the edges of the basin, diffracted waves are generated.
- At t=10s, the diffracted waves, coming from the left and right edges, meet each other at the center of the basin (point B in Fig. 6) and produce large amplifications along the basin. According to Fig. 7, they cross the whole basin 3 times showing the influence of damping on the propagation of the Ricker wavelet.
- At the center of the whole basin (point B), we can estimate the delay between the different arrival times of the laterally propagating waves. The 1D response at the center gave $t_1$=5s, afterwards we have $t_2$=14s and $t_3$=33s. We can then estimate the time delay $\tau$: $\tau=t_2-t_1$=9s and $t_3-t_2$=19s, which is approximately 2$\tau$. The mean velocity of the lateral waves can then be estimated as follows: $V_{mean}$=W/$\tau$=500m/s, with W=4500m the valley width. These lateral waves contribute to the longer duration of the ground motion.

To better understand the influence of laterally propagating waves generated at the basin edges, a Ricker wavelet of larger central frequency is now considered (f=0.635Hz). Furthermore, two different geological structures are investigated: a homogeneous basin of mean shear wave velocity 530m/s and the previous 7 layers heterogeneous model. The ground motion all along the basin surface is computed for the 0.635Hz Ricker wavelet. This frequency value corresponds to the third mode of the valley (Fig. 5).

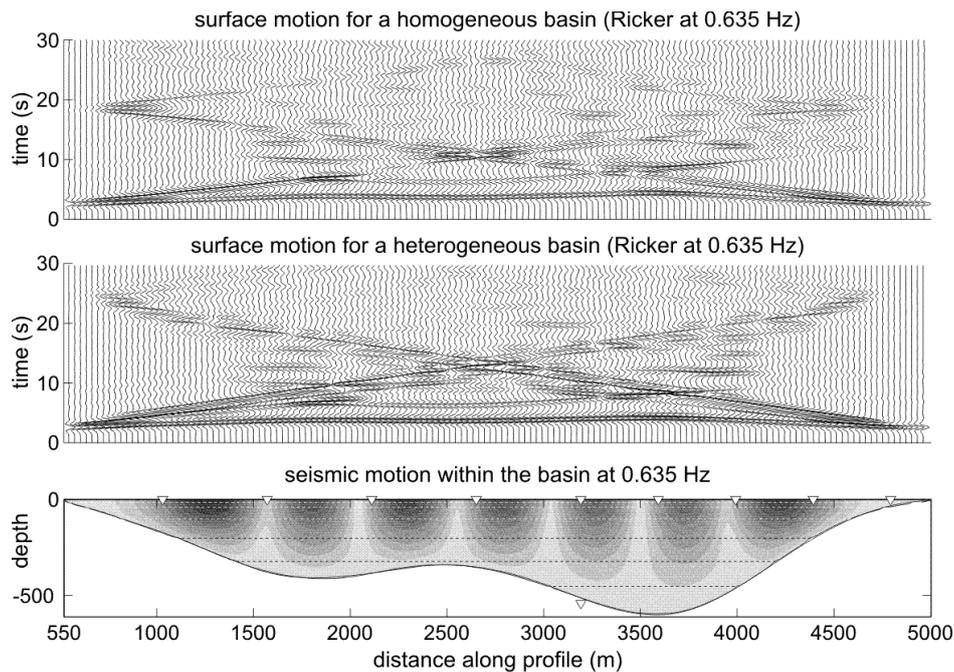

**Fig. 8.** *Synthetic seismogram for a vertically incident SH-wave (Ricker signal at 0.635Hz).*
*Top: homogeneous basin (Vs=530m/s); middle: 7 layers model.*
*Bottom: ground motion amplification within the basin at a given frequency (f=0.635Hz).*

In Fig. 8, the ground motion for the homogeneous basin model (top) is compared to the heterogeneous 7 layers case (middle). As suggested by Luzon et al. [51] and Semblat et al. [13], the heterogeneous 7 layers model (middle) leads to a larger amount of laterally propagating waves than the homogeneous basin model (top). As also shown in Fig. 8 (bottom), the spectral amplification is strong for various areas along the free surface (maximum spectral amplification reaching 21.5 in the darkest area). However, for larger frequencies, one may have significant amplification within the basin (first in its deepest part).

4.1.4. Amplification of higher frequency components.
To investigate the amplification of higher frequency components and thus the influence of the surficial soft layers, the seismic motion within the basin is displayed in Fig. 9 for frequencies 0.7, 0.9, 1.1 and 1.3Hz. The related maximum spectral amplifications (darkest areas) are as follows: 6.5, 5.7, 11.1 and 8.6. At 0.9Hz, the in-depth seismic motion is amplified at a significant level. At 1.1Hz, the in-depth amplification is obvious in both left and right part of the basin. For frequencies above 1.3Hz, the amplification of the surface motion mainly concerns the soft surficial layers. It clearly shows the influence of the velocity profile on the amplification process. The analysis of the focusing phenomena is also very interesting for the deepest left and right parts of the basin as well as for the central shallower area.

Such higher frequency simulations cannot be performed in the 3D case with the classical Boundary Element Method yet. The recent accelerated BEM formulations [43-44,47] may reach such higher frequencies in 3D simulations [52] but should be improved to deal with large velocity contrasts.

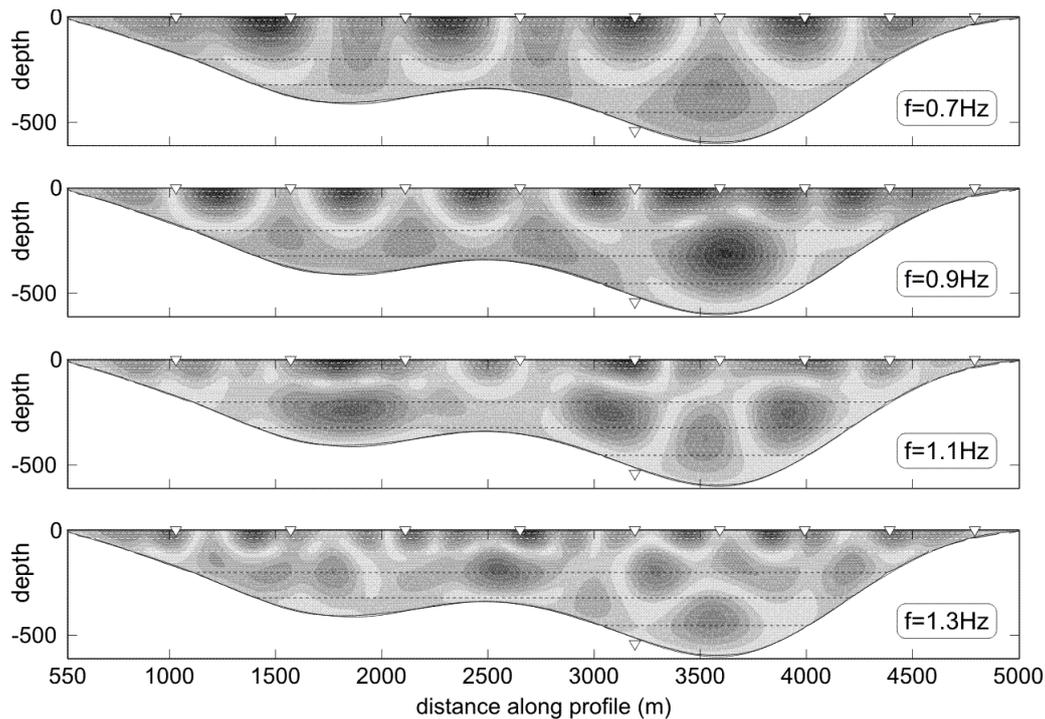

**Fig. 9.** *Seismic motion within the basin at various frequencies; the maximum spectral amplifications (darkest areas) are 6.5, 5.7, 11.1 and 8.6 respectively.*

### *4.2. SH point source*

As suggested in the benchmark [21-22] (Fig. 2), we also analyzed the case of an earthquake due to a fault located to the east, near Grenoble area (events W1 & S1 of the numerical benchmark). This fault is perpendicular to the profile: the projection of the fault on the 2D profile is a line. We approximated this line with 3 point sources corresponding to the top, the middle and the bottom (s=10645m; $z_1$=-750m, $z_2$=-3000m, $z_3$=-5250m). The epicentral distance *d* was estimated with respect to the point B (Fig 6) at the center of the profile: the source is then projected along the profile at a distance *d* from the point B. Such 2D point sources correspond to sources along horizontal lines in 3D. A 2.5D BEM approach [14] may be needed to better model realistic 3D point sources.

4.2.1. Amplification in the frequency domain
Considering the transfer functions for the point source excitation (Fig. 5, bottom right), the maximum amplification reaches 22 for some receivers. The same fundamental frequency value as in the plane wave case is recovered (f=0.33Hz). If the directivity effects are also taken into account, the point source results are rather close to those of the plane wave case with the incidence angle θ=135° (Fig. 5, top right).

4.2.2. Time domain amplification of the seismic motion.
As in the plane wave case, we considered a Ricker wavelet of central frequency f=0.325Hz. It corresponds to the fundamental frequency of the basin (Fig. 5). The ground motion is then computed in time domain all along the basin surface (Fig. 10).

A strong motion amplification can be observed on the right part of the basin: scattered waves are mainly generated on the right edge of the basin and the directivity effect is found to be strong. The laterally propagating waves coming from the right edge of the basin appear clearly and cross the whole basin 3 times in the time interval considered (60s). A much weaker wave field is generated at the left edge of the basin and can be observed on the right part of the valley around time t=25 s.

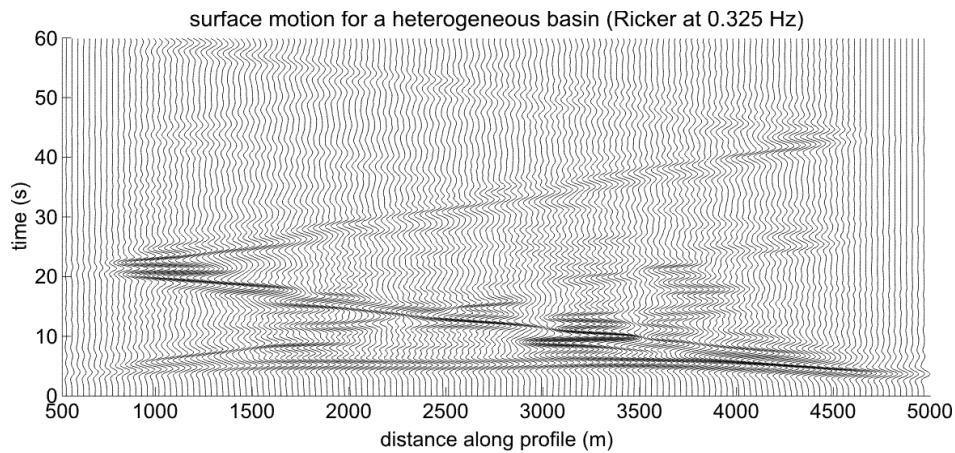

**Fig. 10.** *Synthetic seismogram for point source SH-waves (Ricker signal at 0.325Hz).*

## 5. Incident plane SV-wave
### 5.1. Amplification in the frequency domain

We now estimate the amplification on the basin surface for a plane SV wave with a vertical and oblique incidence. For a vertical incidence, the horizontal motion component (Fig. 11, top left) reaches a maximum amplification value of 12 in some areas. The corresponding fundamental frequency is 0.38Hz. For the vertical incidence, the amplification of the vertical component (Fig. 11, top right) is much lower since the maximum level is nearly 6. For the oblique incidence (Fig. 11, bottom), the first peak of the transfer function for the horizontal component is associated with a larger frequency value than in the SH case (f=0.7Hz instead of f=0.325Hz). Concerning the amplification level, it is generally lower in the SV case (Fig. 11) than in the SH case (Fig. 5).

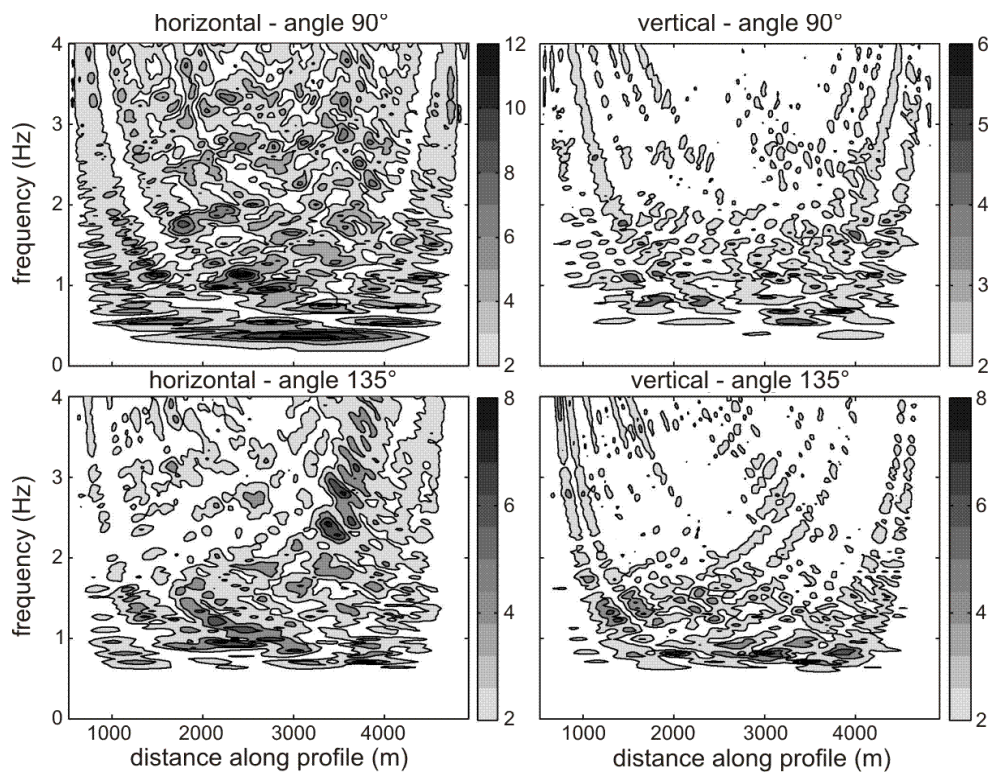

**Fig. 11.** *Transfer functions of the horizontal (left) and vertical (right) motion components for a vertical (top) and oblique (bottom) plane SV wave.*

## 5.2. *Time domain amplification of the seismic motion.*

In this section, as suggested in the numerical benchmark, the main goal is to perform realistic seismic motion computations. The plane SV wave transfer functions are considered for a vertical incidence angle. A synthetic signal is considered using the procedure proposed by Pousse et al. [53] as a seismic input for a Magnitude 6 earthquake.

In Fig. 12, the horizontal component of the velocities for 10 receivers along the profile is displayed. The PGV is 0.08m/s for the input and reaches PGV=0.22m/s at receiver R6 (located 3600m from the left edge of the basin). The time duration of the signal is equal to 3 times that of the input signal. To model strong seismic motions, it may be necessary to consider nonlinear constitutive laws in the surficial soil layers [27-29]. Such parameters were available in the benchmark for a few 1D soil profiles.

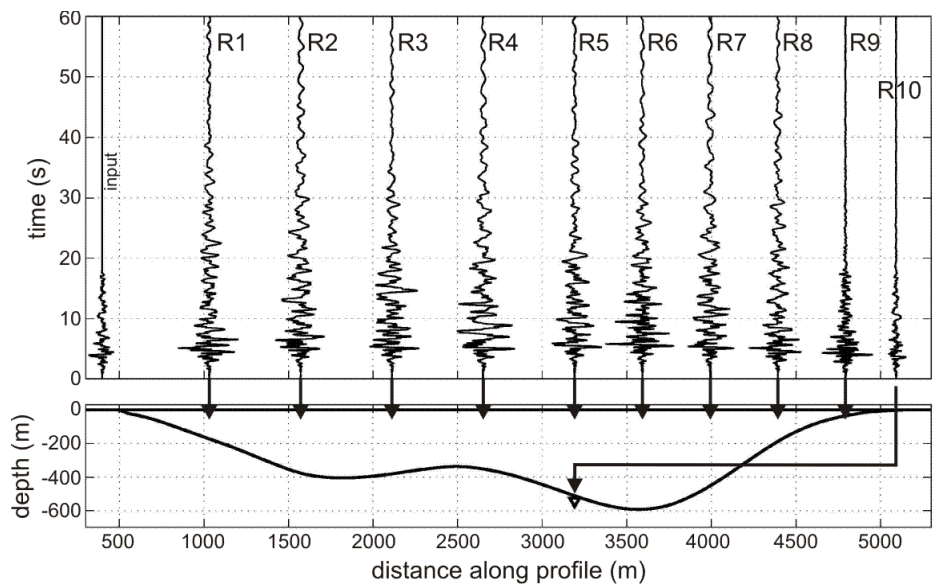

**Fig. 12.** *Horizontal component of the velocity for a vertical SV wave at the 10 receivers given by the numerical Benchmark (Mg. 6 earthquake at a hypocentral distance of 20km of the basin).*

## 6. Conclusion

Strong site effects have been characterized for an alpine valley (Grenoble basin, French Alps) having a strong velocity gradient. Since the classical BEM is limited to weakly heterogeneous media and leads to a large computational burden in 3D, the 2D cross-section proposed in the ESG benchmark was chosen [21-22]. In this 2D model, the computed spectral amplification reaches a maximum value of 20. The amplification process is strong near the free surface at moderate frequencies. As shown by other numerical results [21-22], the amplification within the basin itself can be significant for larger frequencies (shorter wavelengthes).

The results detailed herein mainly show the specific 2D phenomena governing site effects for a deep basin with such a strong velocity gradient: low fundamental frequency, strong focusing and directivity effects, etc. In the time domain analysis, such basin properties produce multiple lateral waves on the edges that are strongly trapped in the deep valley.

However, to model the complex 3D geometry of the Grenoble basin, as also proposed in the ESG benchmark, the capabilities of the classical Boundary Element Method considered herein are not sufficient. The recent advances in the field of fast BEM approaches [43] lead to a much smaller computational burden and will allow to deal with such complex 3D simulations in the near future. A first attempt was recently made through a preconditioned 3D fast BEM model of the Grenoble basin considering a large velocity contrast but a homogeneous undamped basin [52]. The Fast Multipole BEM is currently being generalized to viscoelastodynamics [47]. Nevertheless, it should be further improved to model deep 3D basins with strong velocity gradients as proposed in the ESG benchmark. Furthermore, considering the city of Grenoble at the basin surface, one may try to investigate the influence of the buildings on "free-field" site effects since it may be significant in densely urbanized areas [54-56].